# Usability testing:

# some current practices and research questions

J.M. Christian Bastien

Université Paul Verlaine – Metz, Laboratoire Lorrain de Psychologie (2LP, EA 4165), Équipe Transdisciplinaire sur l'Interaction et la Cognition (ETIC), UFR Sciences Humaines & Arts, BP 30309, Île du Saulcy, 57006 Metz, France

# Abstract

The aim of this paper is to review some work conducted in the field of user testing that aims at specifying or clarifying the test procedures and at defining and developing tools to help conduct user tests. The topics that have been selected were considered relevant for evaluating applications in the field of medical and health care informatics. These topics are: the number of participants that should take part in a user test, the test procedure, remote usability evaluation, usability testing tools, and evaluating mobile applications.

## Keywords

# Introduction

As indicated by Saintfort, Jacko and Booske [1], health care informatics "comprise the generation, development, application, and testing of information and communication principles, techniques, theories, and technologies to improve the delivery of health care with a focus on the patient/consumer, the provider, and, more important, the patient-provider interaction" (p. 811). In addition to improving the delivery of health care, these technologies are meant to increase patient safety by reducing medical errors. The field of health care informatics thus defined, comprises a wide variety of technologies and applications that may be used, in different contexts by different kinds of people having their own objectives. The patients/consumers interact with different online services and web sites in order to find medical information to understand their health and make health-related decisions for themselves. They also use the Internet for discussing health matters with other people. Hand-held devices and smart phones are used to engage in more healthy behaviors (i.e., to access to tailored nutrition information [2], to boost physical activity [3], or to manage chronic diseases such as asthma, diabetes [4]).

Health care providers already use medical devices [1, 5] that incorporate human-computer interfaces: infusion pumps, bar coding medication systems (BCMA), computerized physician order entry (CPOE), clinical decision support systems (CDSS), monitors, surgical robots, electronic medical records (EMR), radiology systems, etc.

To fulfill the goals of improving the delivery of health care and increasing patient safety, these technologies must demonstrated as not being error prone. Unfortunately, as indicated by Beuscart-Zéphir *et al*. [5] a number of cases have been documented that show that medical devices (i.e., infusions pumps [6], handheld e-prescribing application [7], CPOE [8-11], etc.) are in fact error prone.

Usability evaluation is one way of ensuring that interactive systems are adapted to the users, their tasks and that there are no negative outcomes of their usage. Usability evaluation is a fundamental step in the user centered design process [12] of any interactive system be it a software, a web site or any information and communication technology or service. The goal of a usability evaluation is to assess the degree to which a system is effective (i.e., how well the system's performances meet the tasks for which it was designed), efficient (i.e., how much resources such as time or effort is required to use the system in order to achieve tasks for which the system was design), and favors positive attitudes and responses from the intended users [13].

The three standard approaches for evaluating user interfaces are *Inspection-*, *User-*, and *Model-Based Evaluations*. Although these evaluation methods were not originally developed for medical interactive systems, their use in the health care settings as increased during the last ten years. The first two approaches are the most widely used by usability practitioners [14, 15] and have been extensively documented. The model-based approaches are considered limited or immature, expensive to apply and their use is largely restricted to research teams [16].

When one look at the books and articles on usability testing, one gets the impression that everything has been said, and that no research questions are left unanswered. However, the standard usability test as is currently applied to test most computer software applications show some limitations or at least raises some questions when applied to some specific domains in which users surf the Internet or use mobile devices. The purpose of this article is not to present the three approaches in details. It neither aims at providing a framework for evaluating healthcare applications. In this respect, the reader should look at the book chapter by Sainfort, Jacko and Booske [1]. The aim of this paper is to review some work conducted in the field of user testing that aims at specifying or clarifying the test procedures and at defining and developing tools to help conduct user tests. This review is not exhaustive. The topics that have been selected were considered relevant for evaluating applications in the field of medical and health care informatics mentioned previously. These topics are: the number of participants that should take part in a user test, the test procedure, usability testing tools, remote usability evaluation and user testing mobile applications.

## User-based evaluation

User-based evaluations are usability evaluation methods in which users directly participate. Users are invited to do typical tasks with a product, or simply asked to explore it freely, while their behaviors are observed and recorded in order to identify design flaws that cause user errors or difficulties. During these observations, the time required to complete a task, task-completion rates, and number and types of errors, are recorded. Once design flaws have been identified, design recommendations are proposed to improve the ergonomic quality of the product.

The user test or empirical usability test is well documented [17-22]. The implementation of a user test generally goes through a certain number of steps such as:

- the definition of the test objectives,
- the qualification and recruitment of tests participants,
- the selection of tasks participants will have to realize,
- the creation and description of the task scenarios,
- the choice of the measures that will be made as well as the way data will be recorded,
- the preparation of the test materials and of the test environment (the usability laboratory),

- the choice of the tester, and the design of the test protocol per se (instructions, design protocol, etc.),
- the design and/or the selection of satisfaction questionnaires, the data analyses procedures,
- and finally the presentation and communication of the test results.

Some of these steps, as we will see, raise some questions that are still difficult to answer while others are still waiting for the development of useful and usable tools. The topics that will be addressed are: the number of participants one has to recruit for conducting a user test, the test procedure, conducting user test remotely, the tools available and needed to conduct usability tests, and the evaluation of mobile applications and services.

### *How many users do we have to test?*

Deciding how many users to recruit has both practical/economic and scientific implications. When inviting users to participate in a user test, the aim is to find the most design flaws a user interface may have, at the lowest cost (cost of participants, cost of observers, cost of laboratory facilities, and limited time to obtain data to provide to developers in a timely fashion [23]). In this respect, one must ensure, based on experimental evidence, that the number of tests participants will allow a complete evaluation of the interface being evaluated and that no superfluous users will be recruited. This point has been studied since the nineties and has not find a final answer yet [23, 24].

In the nineties, it was said that with 4 or 5 participants, 80 to 85% of the usability problems of an interface could be uncovered [25-27]. However, Spool and Schroeder [28] published the results of a large-scale usability evaluation in which they concluded that the Web sites they studied would need considerably more than five users to find 85% of the usability problems. In this study, 49 participants took part in a user test in which they had to describe an item they wanted and buy it. This task was repeated on 4 different Web sites. The results indicate that with the first 5 participants only 35% of the usability problems were uncovered. In addition, serious problems that prevented intended purchases were only found with the 13th and 15th participant. The type of interface as well as the tasks participants had to do may explain the differences observed. One must recall that the results obtained in previous studies came from the study of mainframe applications, videotext and PC applications. In the case of Web sites, users have to make many personal choices [28]. This question of the number of users to test is far from being solved and requires further research. These issues are of particular importance since the evolution towards eHealth will involve the use of Internet for Web sites for publishing consumer/patient content, and providing health care professionals with Web-based applications.

## *The test procedure*

Most of the usability test sessions are run with a single test participant. However, in some cases, test sessions may be conducted with two participants working together. Both cases have advantages and drawbacks [18, 20]. Inviting two participants to take part in a test session has sometimes been used to alleviate the difficulties or feelings of unease some participants may experience in individual sessions. In paired-user testing, or codiscovery evaluation, participants are invited to accomplish together some tasks on the same computer. In this context, we observe an increase in the number of utterances participants spontaneously make and also an increase in the number of discussions and justifications on how to achieve the tasks the participants engage in. In these situations, interactions between the tests participants take precedence over the interactions with the evaluator. O'Malley, Draper and Riley [29] were the first to describe this procedure which they called "Constructive interaction". This procedure is not well documented and the rare publications on this topic present few quantitative data. This is the case for Wildman [30] who only describe the procedure and Westerink, Rankin, Majoor and Moore [31] who use this protocol with adolescents in the evaluation of computer games. In spite of the few data on this protocol Wilson and Blostein [32] provide a list of pros and cons. On the positive side we find that the paired-user testing is good for early design phase, promotes a natural interaction style, produces more comments than think-aloud sessions, is easier for the experimenter, is a good method for applications where people work together and is more fun for both participants. However, participants' different learning, verbal, cultural or hierarchical styles may affect feedback. Careful candidate screening is thus needed. In addition, more participants are necessary and the data analysis is harder. In a study we conduct on paired-user testing [33], thirty-two participants had to perform 8 tasks designed to allow the evaluation of interactive television services. For half of the participants (16) the test session was conducted individually: for the other half, the test session employed pairs of users (8x2 participants) performing the tasks together. The main results indicated that task completion times did not differ statistically between groups and that paired-user testing involved better success rates. Individual sessions allowed the identification of more usability problems while paired-user sessions allowed a better understanding of the difficulties users encountered.

### *Remote usability evaluation*

Most of the time, usability evaluations are conducted in a usability laboratory. People that were recruited are invited to come to the test facilities consisting of a test room, where the participants will accomplish specific tasks, an observation room and the "recording" room. A usability laboratory may contain complex and sophisticated audio/visual recordings and analysis facilities. In this context, test sessions are conducted individually. Although this situation has advantages it also has drawbacks, as we will see.

Remote usability evaluation refers to a situation in which the evaluators and the test participants are not in the same room or location. Two approaches to remote usability evaluation have been developed: synchronous and asynchronous. Each approach uses specific tools. In the synchronous approach, a facilitator and the evaluators collect the data and manage the evaluation session in real time with a participant who is remote (the participant may be at home, at work or in another room). The evaluation may require video conferencing applications or remote applications sharing tools that allow to share computer screens so as to allow the evaluator to see what is happening on the user's screen (tools such as WebEx (http://www.webex.com), Microsoft NetMeeting or Lotus Sametime (http://www-306.ibm.com/software/lotus/sametime/)) [34, 35]. In contrast, with asynchronous methods, observers do not have access to the data in real time, and there is no facilitator interacting with the user during data collection. Asynchronous methods also include automated approaches, whereby users' click streams are collected automatically (e.g., WebQuilt). The key advantage this technique offers is that many more test users can participate (in parallel), with little or no incremental cost per participant. For conducting these asynchronous tests, different strategies have been proposed. One strategy is to ask test participants to download and use an instrumented browser that will capture the users' click streams as well as screen shots, and transmit those data to the evaluator's host site for analysis (an example of this kind of browser is ErgoBrowser, http://www.ergolabs.com/resources.htm). Another approach consists in using a proxy. The test participants are invited to go to a specific Web site and then to follow instructions. They are then brought to the Web site under evaluation. The users' behaviors are captured, aggregated and visualized to show the web pages people explored. The visualization also shows the most common paths taken through the website for a given task, as well as the optimal path for that task as implemented by the designer. An example of this kind of approach is WebQuilt [36] and the work by Atterer, Wnuk and Schmidt [37].

The asynchronous approach does not allow for observational data and recordings of spontaneous verbalizations during the remote test sessions. The qualitative data can only be recorded through post-test questionnaires or self-report forms. However, the asynchronous approach allows the recording of large groups of users as we said.

The synchronous approach is favored by some authors [34] because it is analogous to laboratory testing and because it allows the capture of qualitative data. In comparison to the laboratory user test, the synchronous remote testing is cost effective, especially for travel expenses when participants are recruited in different region in a given country. However, the costs associated with this approach may in some cases be quite similar to those of the laboratory testing (for the recruitment for instance). Two other reasons for preferring the remote synchronous approach to traditional user testing is the freedom from facilities (especially when the product or software can be distributed electronically or when testing a Web site) and time saving. However synchronous remote testing can be perceived as more intrusive than traditional laboratory user testing.

The question one must answer before choosing one approach over the other is how they compare to the traditional user testing in terms of usability problems uncovered. One study [38] has demonstrated that the synchronous remote testing yields comparable results to a traditional user test of the same application. Tullis *et al.* [39] present results that show high correlations between laboratory and remote tests for the task completion data and the task time data. The most critical usability issues with the web sites were identified by both techniques, although each technique also uniquely uncovered other issues. In general, the results indicate that both the laboratory and remote tests capture very similar information about the usability of a site. Another study by West and Lehman [40] was conducted to evaluate a method for usability testing with an automated data collection system. They found it to be an effective alternative to a laboratory-based test. The remote testing results presented only minor differences in comparison to laboratory evaluation. These results are consistent with those of Tullis *et al.* [39]. In a recent study [41] three methods for remote usability testing and a traditional laboratory based think-aloud method were compared. The three remote methods were a remote synchronous condition, where testing was conducted in real time but the usability evaluator was separated spatially from the test participants, and two remote asynchronous conditions, where the usability evaluator and the test subjects were separated both spatially and temporally. The results showed that the remote synchronous method was equivalent to the traditional laboratory method. The asynchronous methods were considerably more time consuming for the test subjects and identify fewer usability problems.

As can be seen from these studies, work is still needed on these aspects. The type of Web sites studied, the instructions given to the test participants, the coding of the participants' behaviors, the analysis of the data as well as the procedures may explain the differences observed between these studies.

### *User testing tools for the usability specialist*

Most of the time, user test sessions are audio and video recorded. These recordings are than viewed and coded with a behavior grid. The evaluator determines the frequency, duration of all the behaviors that can indicate user problems or difficulties as well as performance measures such as time to finish a task, time spent recovering from errors, number of wrong icon choices, observations of frustrations, of confusion and satisfaction, etc. This coding, to be precise, necessitates specific hardwares and softwares, and is very time consuming. To increase the efficiency of user tests, software tools are being developed. Some of these tools allow the evaluator to control video recordings in order to identify precisely the beginning and the end of a behavior and the duration of tasks. In some cases, these professional softwares provide the evaluators with descriptive statistics on the behaviors observed (frequencies, mean duration, total duration, etc.) as well as behavior patterns (e.g., The Observer and Theme from Noldus, http://www.noldus.com) [42]. The use of such applications reduces significantly the time dedicated to the coding of video recordings. Morae

(http://www.techsmith.com/morae.asp) is another software used for recording and logging user interactions. It is particularly suited for user testing Web sites. It allows the recording of user interactions with a Web site or application, including desktop activity, audio, camera video and a complete chronicle of system events, all synchronized into a single file. The software lets the evaluators analyze and visualize the data and select video sequences for highlighting specific interactions. All the events captured can be exported to statistical sofwares.

We have seen in the preceding sections that the automatic recording of users' actions was important for several reasons and that tools were developed for this purpose. In fact, the automatic recording of users' interaction has been addressed by some researchers and not only for user testing the Web. We should say that this problem was addressed long before we studied users interacting with the web. We were first concerned with users interacting with menus, menu options and dialogue boxes. Some early tools [43] allowed the graphical comparisons of novices and experts behaviors. But the problem with the automatic recording of user interactions is the amount of data recorded and the degree of granularity of the analyses. More recent tools (e.g., KALDI) [44] allow both the recordings of users' behaviors and the recording of the interface elements displayed. Such a tool allows to represent graphically user actions and to display them according to different level of abstraction (elementary events, tasks, etc.).

As indicated by Ivory and Hearst [45], usability evaluation can be expensive in terms of time and human resources. Automation is therefore a promising way to augment existing approaches by reducing the cost of usability evaluation (by eliminating the need for manual logging of user events), by increasing consistency of the errors uncovered, by increasing the coverage of evaluated features, etc. However, tools currently available, although representing a valuable help on certain aspects of the user test (date capture, data analysis, data representation) are incomplete. The available solutions are not yet integrated.

*Evaluating mobile application*

Another aspect, which is gaining importance, is the evaluation of mobile devices (e.g., telephones, smartphones, PDA), applications and services. As indicated by Sainfort, Jacko and Booske [1] wireless, handheld and mobile technologies will increasingly become an important part of healthcare's information technologies. With these technologies and especially mobile technologies, we find all the problems listed previously plus the fact that usability evaluation of such services should take into account the specific context, i.e., "mobility". This aspect cannot be studied, by definition, in the laboratory even if some researcher have reproduced some characteristics of mobility by having users walk on treadmills to simulate walking conditions. As indicated by Schusteritsch, Wei and LaRosa [46], "for most mobile usability studies, enabling natural interaction with the device can be more challenging than in a desktop-based environment because mobile phones come in a diverse range of shapes and run a variety of operating systems. Depending on study goals, an observation system can be customized to a specific phone model or it may need to be flexible to accommodate a variety of phones. Radically different input systems such as scroll wheels, custom menu buttons, and styluses may have to be supported, and in many situations, the ability for users to hold the mobile device naturally can be critical to capture unbiased interaction patterns." But this is only one aspect of mobility. When we want to evaluate how users will use the services and applications in a natural context, another approach must be put in place. The use of diaries [47], log files, traces and periodic interviews may be used. Diary studies are used to capture activities that occur in real environments with a technology, application or service. In these studies, participants are asked to record particular activities as they occur or to record afterwards the behaviors they were engaged in, on a paper diary.

These diaries can be highly structured, with specific pre-defined categories of activities to be checked off and later counted, such as the number of communications over the course of a given period. They can also be unstructured, with spaces for recording, time-stamping, and describing activity. The problem with diaries is that users sometimes forget to fill in the information or fill it in after a period of time thus relying on memory… However, diaries can be used to complement techniques such as logs and traces. Although the diary studies have been used for a while in applied psychology, it has not been used frequently in usability studies in spite of their potential usefulness (see Palen and Salzman [48] and Rieman [47] for examples of diary studies). However, given the importance mobile applications and service acquire, some annual conferences on mobility and ubiquity (UbiMob) or on human-computer interaction with mobile devices and services (MobileHCI) encourage communications on usability evaluation methods. New and validated approaches and methods should be available in a near future.

## Conclusion

The aim of this paper was to review some work conducted in the field of user testing that aims at specifying or clarifying the test procedures and at defining and developing tools to help conduct user tests. This paper was also aimed at showing to the reader how complicated user testing could be in specific situations. Although the review was far from exhaustive, we hoped it gave an idea of the work conducted as well as the work needed to develop valid usability evaluation methods such as user testing.